\documentclass[]{aa}  

\usepackage{graphicx}
\usepackage{txfonts}
\newcommand{\ebv}{$E(B-V)$ }

\newcommand{\ha}{\hbox{H$\alpha$}}
\newcommand{\hb}{\hbox{H$\beta$}}

\newcommand{\gsim}{\lower.5ex\hbox{$\; \buildrel > \over \sim \;$}}
\newcommand{\lsim}{\lower.5ex\hbox{$\; \buildrel < \over \sim \;$}}

\newcommand{\oiii}{\hbox{[O\,{\sc iii}]}}
\newcommand{\nii}{\hbox{[N\,{\sc ii}]}}

\begin{document} 

        \title{Spatially resolved mass-metallicity relation at $z\sim$0.26 from the MUSE-Wide Survey}


        \author{Yao Yao
                \inst{1}\fnmsep\inst{2}
                \and
                Guangwen Chen\inst{1}\fnmsep\inst{2}
                \and
                Haiyang Liu\inst{1}\fnmsep\inst{2}
                \and
                Xinkai Chen\inst{1}\fnmsep\inst{2}
                \and
                Zesen Lin\inst{1}\fnmsep\inst{2}
                \and
                Hong-Xin Zhang\inst{1}\fnmsep\inst{2}
                \and
                Yulong Gao\inst{3}\fnmsep\inst{4}
                \and
                Xu Kong\inst{1}\fnmsep\inst{2}\fnmsep\inst{5}
        }
        
                \institute{Department of Astronomy, University of Science and Technology of China, Hefei 230026, China
                \and
                School of Astronomy and Space Sciences, University of Science and Technology of China, Hefei, 230026, China
                \and
                School of Astronomy and Space Science, Nanjing University, Nanjing 210093, China
                \and
                Key Laboratory of Modern Astronomy and Astrophysics (Nanjing University), Ministry of Education, Nanjing 210093, China
                \and
                Frontiers Science Center for Planetary Exploration and Emerging Technologies, University of Science and Technology of China, Hefei, Anhui, 230026, China
        }

        \date{Received 13 January 2022 / Accepted 24 March 2022}

 
        \abstract
        {}
        {Galaxies in the local universe have a spatially resolved star-forming main sequence (rSFMS) and mass-metallicity relation (rMZR). We know that the global mass-metallicity relation (MZR) results from the integral of the rMZR, and it evolves with redshift. However, the evolution of the rMZR with redshift is still unclear because the spatial resolution and signal-to-noise ratio are low. Currently, too few observations beyond the local universe are available, and only simulations can reproduce the evolution of the rMZR with redshift.}
        {We selected ten emission-line galaxies with an average redshift of $z\sim 0.26$ from the MUSE-Wide DR1. We obtained the spatially resolved star formation rate (SFR) and metallicity from integral field spectroscopy (IFS), as well as the stellar mass surface density from 3D-HST photometry. We derived the rSFMS and rMZR at $z\sim 0.26$ and compared them with those of local galaxies.}
        {We find that the rSFMS of galaxies at $z\sim 0.26$ has a slope of $\sim$0.771. The rMZR exists at $z\sim 0.26$, showing a similar shape to that of the local universe, but a lower average metallicity that is about $\sim$0.11 dex lower than the local metallicity. In addition, we also study the spatially resolved fundamental metallicity relation (rFMR) of these galaxies. However, there is no obvious evidence that an rFMR exists at $z\sim$0.26, and it is not an extension of rMZR at a high SFR.}
        {Similar to their global versions, the rSFMS and rMZR of galaxies also evolve with redshift. At fixed stellar mass, galaxies at higher redshift show a higher SFR and lower metallicity. These suggest that the evolution of the global galaxy properties with redshift may result from integrating the evolution of the spatially resolved galaxy properties. }


        \maketitle
        \titlerunning{rMZR at $z\sim$0.26}
%

\section{Introduction} \label{sec:introduction}

We can understand the evolution of galaxies by analyzing their gas-phase chemical abundance. The inflow and outflow of gas, the stellar wind, and supernova explosions play a critical role in shaping the gas-phase metallicity ($Z$) of the galaxies. These processes are closely related to other properties of galaxies, the most important of which is the stellar mass ($M_*$). Therefore, the mass-metallicity relation (MZR) should naturally appear \citep{Lequeux1979}. At the same time, the inflow and outflow of gas can regulate the enrichment process of metals. The inflowing gas dilutes the metal and increases the star formation rate (SFR), and the feedback from the star formation and active galactic nucleus (AGN) will blow away the metal that has formed. Some studies \citep{LaraLopez2010, Mannucci2010} have found that the three parameters $M_*$, $Z$, and SFR can form a closer relation, called fundamental metallicity relation (FMR). In this relation, a galaxy with a higher
SFR at a given $M_*$ tends to have lower $Z$, which does not evolve with the redshift.

On the one hand, as the redshift survey continues to deepen, the MZR has been observed at different redshifts in recent decades, from the local universe to $z\sim$3.5. It has been discovered that the MZR will evolve with the redshift \citep[e.g.,][]{Tremonti2004,Maiolino2008,Zahid2014,Gao2018,Huang2019,Sanders2021}. This evolution has many reasons. For example, the galaxies in the local universe have longer histories, which means that they have experienced a longer period of gas consumption and metal enrichment than high-redshift galaxies. \cite{Yabe2015} hypothesized that galaxies at high redshift have more gas inflow and outflow, while \cite{Lian2018} conjectured that galaxies at high redshift have more metal outflow and a steeper slope of the initial mass function (IMF). \cite{Ma2016} used high-resolution cosmological zoom-in simulations and found that the evolution of the MZR is associated with the evolution of the stellar and gas mass fractions at different redshifts.

On the other hand, spatially resolved spectroscopy of local galaxies became feasible, and resolved-to-scale relations can be detected \citep{Sanchez2020}. The corresponding surface densities replace all extensive quantities in the resolved scaling relation. Integral field spectroscopy (IFS) surveys, such as CALIFA \citep{Sanchez2012} and MaNGA \citep{Bundy2015} have enabled us to find the local version of global scale relations on subgalactic scales. One of them is the resolved star formation sequence (rSFMS) \citep[e.g.,][]{CanoDiaz2016,Hsieh2017,Pan2018,Jafariyazani2019}, which refers to the relation between the stellar mass surface density ($\Sigma_*$) and the SFR surface density ($\Sigma_{\rm SFR}$). Another is the resolved mass-metallicity relation (rMZR) \citep[e.g.,][]{RosalesOrtega2012,Sanchez2013,BarreraBallesteros2016,Gao2018b}, which refers to the relation between the $\Sigma_*$ and its local metallicity. The rMZR is regarded as a more fundamental relation than the global MZR because it can reproduce the global rate and the metallicity profile along the galactocentric radius well \citep{BarreraBallesteros2016}. Similar to the FMR and rMZR, the spatially resolved FMR (rFMR) can also be developed. However, there is a lack of clear evidence for the existence of the rFMR in studies of $z\sim$0 galaxies \citep[e.g.,][]{Sanchez2013, Gao2018b, Sanchez2020}.

However, due to the low spatial resolution and signal-to-noise ratio (S/N) at high redshift, the evolution of the rMZR and rFMR with redshift is still unclear. Recently, a few works have studied this. \cite{Trayford2019} have used the EAGLE simulation to study it and reported that the normalization of rMZR strongly evolves with redshift. Similar to the evolution of the global MZR, the lower the redshift, the higher the metallicity. They also reported that the shape of the rMZR also evolves strongly when the AGN feedback is taken into account. \cite{Patricio2019} have analyzed three gravitationally lensed galaxies at $z$=0.6, 0.7, and 1. However, they found higher metallicities than in the local universe at the same $\Sigma_*$, possibly caused by the differences in sample selection and metallicity calibration. \cite{Gillman2022} have observed 22 main-sequence galaxies at $z\sim$1.5, but found that the evolution of the rMZR is inconsistent with the evolution of the global MZR. This work uses a sample selected from the IFS survey to focus on the rMZR at higher redshifts and compare it with that of local galaxies. Their rSFMS and metallicity gradient are also studied.

This paper is organized as follows. In Sect. \ref{sec:data} we describe our selection criteria of the sample and the measurements from the data. In Sect. \ref{sec:result} we present our results of the rMZR and other resolved properties and discuss whether our result is an extension of the local rFMR for a high SFR. In Sect. \ref{sec:conclusion} we summarize the main conclusion of this work.
Throughout this paper, we adopt a $\Lambda$-CDM cosmology with $\Omega_\Lambda$= 0.7, $\Omega_m$=0.3, and H$_0$=70 km s$^{-1}$ Mpc$^{-1}$. The IMF we adopted is the \cite{Chabrier2003} IMF.

\section{Data} \label{sec:data}

\subsection{Data source overview} \label{subsec:data_source}
In order to derive the spatially resolved properties of galaxies in the intermediate-redshift universe, we need an IFS data set and multiband photometric images. We selected the spectroscopy of the MUSE-Wide Survey \citep{Urrutia2019} and the photometric data of 3D-HST \citep{Brammer2012,Skelton2014} as the data source.
In addition, we also collected the morphological parameters of galaxies from the Dark Energy Spectroscopic Instrument (DESI) Legacy Imaging Survey \citep{Dey2019}.

\subsubsection{MUSE-Wide DR1} \label{subsubsec:muse}
The MUSE-Wide Survey is a blind 3D spectroscopic survey in the CANDELS/GOODS-South and CANDELS/COSMOS regions. The final survey will cover 100$\times$1 arcmin$^2$ MUSE fields. Each MUSE-Wide pointing has a depth of one hour and hence targets more extreme and more luminous objects over ten times the area of the MUSE-Deep fields \citep{Bacon2017}. The MUSE instrument performed it on the Very Large Telescope (VLT) in the wide-field mode, which provides medium-resolution spectroscopy at a spatial sampling of $0\farcs2$ per spatial pixel and an extended range, covering 4750-9350\AA{} with a resolution of $\sim$2.5\AA. We used the first data release (DR1) \citep{Urrutia2019} of the MUSE-Wide Survey, which covers the first 44 CANDELS/GOODS-S pointings of the 100 final pointings. This data release provides datacubes of the observed 44 fields and a catalog of 1602 emission line sources with redshifts in the range of $0.04 \lesssim z \lesssim 6$. All catalogs and spectra can be accessed at the MUSE-Wide official website\footnote{\url{https://musewide.aip.de}}.

\subsubsection{3D-HST} \label{subsubsec:3dhst}
3D-HST is a Hubble Space Telescope (HST) Treasury program to provide Wide Field Camera 3 (WFC3) and Advanced Camera for Surveys (ACS) grism spectroscopy over four extragalactic fields (AEGIS, COSMOS, GOODS-South, and UDS), augmented with previously obtained data in GOODS-North. In addition to the grism spectroscopy, the project provides reduced WFC3 images in all five fields, extensive multiwavelength photometric catalogs, and catalogs of derived parameters such as redshifts and stellar masses. These ancillary data come from a wide range of other public programs, most notably the CANDELS Multi-Cycle Treasury program \citep{Grogin2011,Koekemoer2011}. We used the GOODS-South field of the first comprehensive photometry release of 3D-HST, dubbed version 4.1, including reduced WFC3 F125W, WFC3 F140W, WFC3 F160W, ACS F435W, ACS F606W, ACS F775W, and ACS F850LP image mosaics, which cover the 44 fields of MUSE-Wide DR1 completely.

\subsubsection{DESI Legacy Imaging Survey} \label{subsubsec:desi}
The DESI Legacy Imaging Surveys produce an inference model catalog of the sky from a set of optical and infrared imaging data, comprising 14,000 deg$^2$ of extragalactic sky visible from the northern hemisphere in three optical ($g, r, z$) and four infrared bands. Since we already have a higher spatial resolution 3D-HST image, we only used the data of the size and ellipticity of the galaxy to select our sample. We used the Sweep catalogs of DESI Legacy Imaging Surveys Data Release 9.

\subsection{Sample selection} \label{subsec:roughselection}
We selected the sample from the MUSE-Wide DR1 emission line catalog, which includes 3057 emission lines of the integrated spectra of 1602 galaxies. 
We used the following rules to select the source. Firstly, the emission lines (\ha, \hb, \oiii$\lambda$5007, and \nii$\lambda$6583) should be detected (the S/N of integrated spectrum is $>$5). Secondly, The galaxy should not be at the edge of the MUSE-Wide field. Thirdly, there should be no AGN feature. Lastly, the effective radius should be R$_{\rm e}>0\farcs6$, and the axis ratio should be $b/a>$0.3 in the rest-frame optical band.

\begin{figure*}[ht!]
        \centering
        \includegraphics[width=1\textwidth]{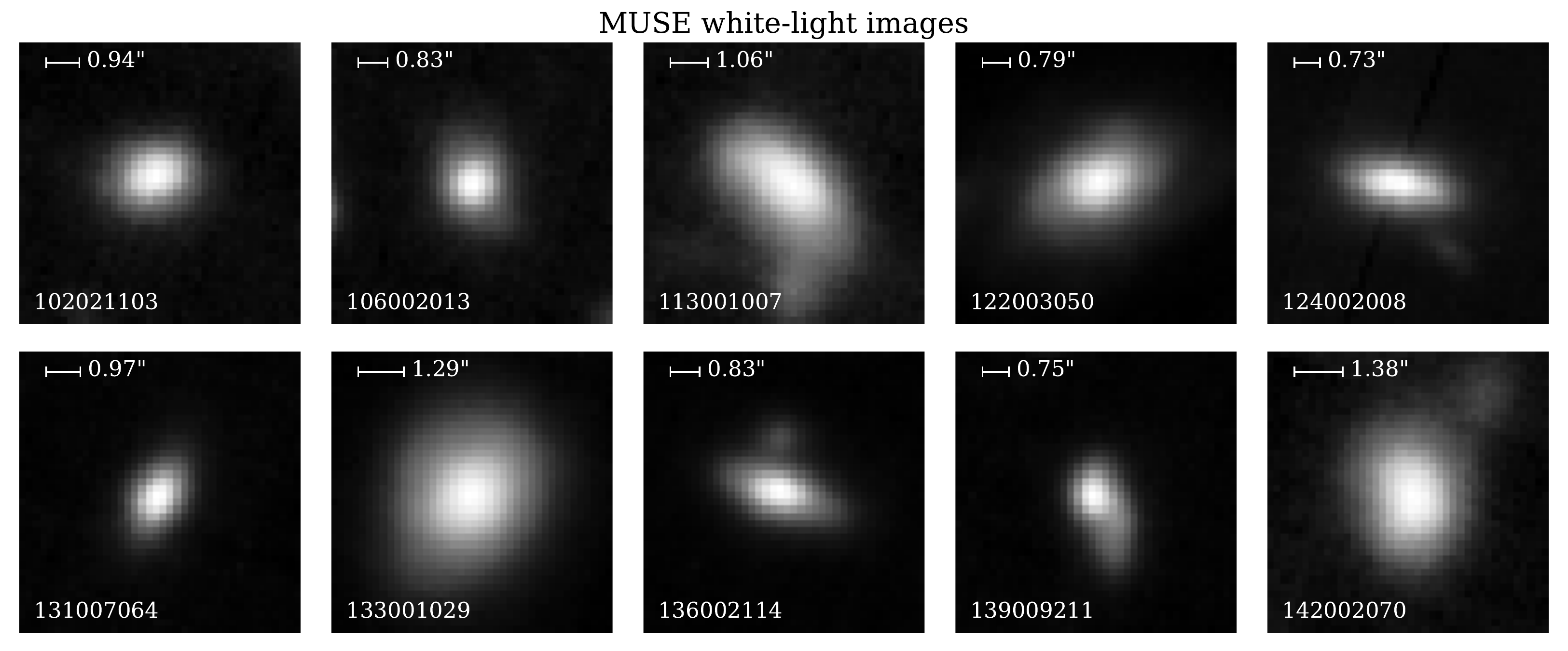}
        \includegraphics[width=1\textwidth]{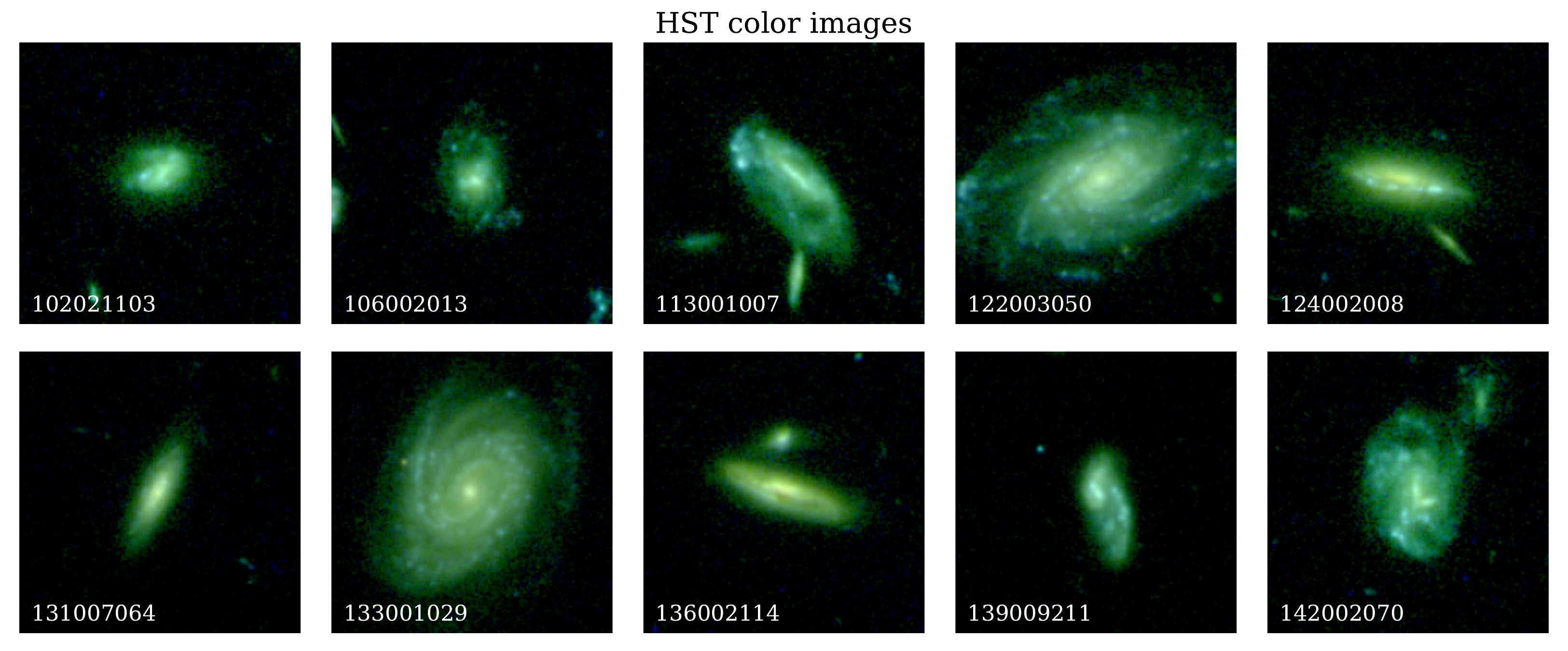}
        \caption{White-light images of MUSE (top two rows), where white error bars represent the FWHM of the PSF.  HST color images (bottom two rows) combined from ACS F435W, F606W, and F775W. }
        \label{fig:image}
\end{figure*}

\begin{figure}[ht!]
        \centering
        \includegraphics[width=0.9\columnwidth]{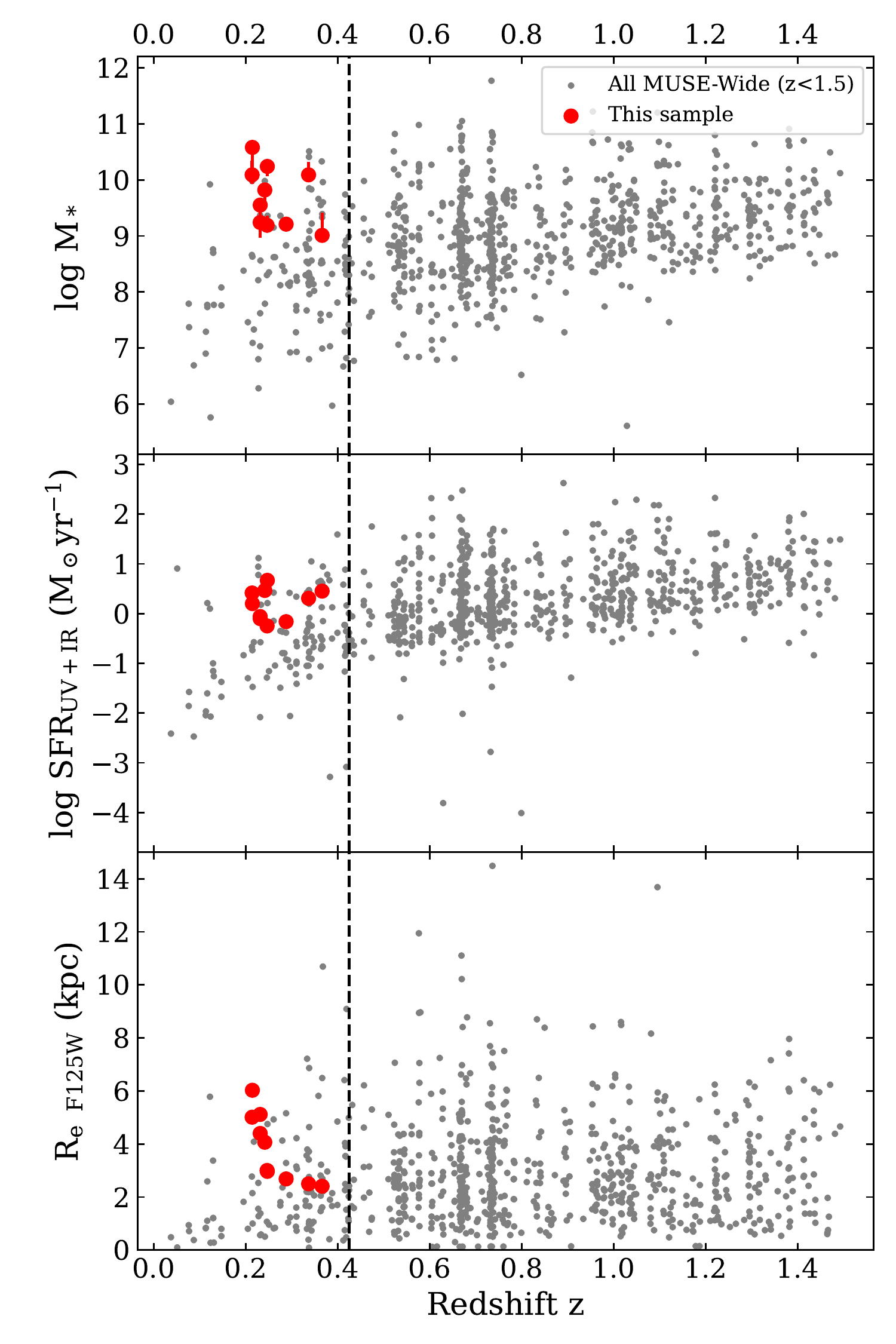}
        \caption{Stellar mass, SFR, and R$_{\rm e}$ of our galaxies (in red) and all low-redshift MUSE-Wide emission line galaxies (in gray). All values are matched from the 3D-HST GOODS-S photometric catalogs. The dashed black line represents the highest redshift at which the \ha{} and \nii{} emission line can be observed ($z\sim$0.42). This is limited by the wavelength range of MUSE.}
        \label{fig:global_info}
\end{figure}

We used the COMMENT column in the emission line catalog to exclude AGN, which are classified by matching the X-ray catalog of the 7Ms Chandra Deep Field South \citep{Luo2017}. We obtained the morphological parameters (R$_{\rm e}$ and $b/a$) by matching the sweep catalog of the DESI Legacy Imaging Survey. After our selection, a total of ten galaxies meet our requirements. Their MUSE white-light images \citep[created from datacubes by summation over the spectral axis;][]{Herenz2017} and HST color images are shown in Fig. \ref{fig:image}. The white error bar in each panel of Fig. \ref{fig:image} indicates the full width at half maximum (FWHM) of the point spread function (PSF) of the MUSE-Wide field. Their detailed information is listed in Table \ref{table:galaxies_info}. The mean redshift of our sample is $\bar{z}\simeq 0.26$. Fig. \ref{fig:global_info} shows the distribution of the stellar mass, SFR, and R$_{\rm e}$ of our galaxies with the distribution of all MUSE-Wide low-redshift galaxies.

\begin{table*}
        \caption{Basic information of the ten galaxies in our sample\label{table:galaxies_info}}
        \centering
        \begin{tabular}{cccccccr}
        \hline\hline
        Unique ID & R.A. & Decl. & Redshift & R$_{\rm e}$ & Sersic n & b/a & Total stellar mass\\
                          & (deg) & (deg) &  & (arcsec) &  &  & (log($M_*$/$M_\odot$))$\hspace{0.5em}$\\
        (1) & (2) & (3) & (4) & (5) & (6) & (7) & (8)$\hspace{2.5em}$\\
        \hline
        \noalign{\smallskip}
        102021103 & 53.0742 & -27.8240 & 0.247 & 0.76$\pm$0.00 & 1.00$\pm$0.01 & 0.662$\pm$0.003 & 9.19$_{-0.08}^{+0.17}\hspace{1em}$ \\
        \noalign{\smallskip}
        106002013 & 53.0803 & -27.8083 & 0.288 & 0.68$\pm$0.01 & 1.87$\pm$0.01 & 0.873$\pm$0.004 & 9.21$_{-0.06}^{+0.11}\hspace{1em}$ \\
        \noalign{\smallskip}
        113001007 & 53.1154 & -27.8271 & 0.232 & 1.79$\pm$0.03 & 1.55$\pm$0.02 & 0.322$\pm$0.003 & 9.24$_{-0.27}^{+0.18}\hspace{1em}$ \\
        \noalign{\smallskip}
        122003050 & 53.1310 & -27.8291 & 0.214 & 1.73$\pm$0.01 & 1.66$\pm$0.01 & 0.578$\pm$0.001 & 10.09$_{-0.16}^{+0.25}\hspace{1em}$ \\
        \noalign{\smallskip}
        124002008 & 53.1617 & -27.8323 & 0.242 & 1.11$\pm$0.00 & 1.34$\pm$0.01 & 0.325$\pm$0.001 & 9.82$_{-0.24}^{+0.12}\hspace{1em}$ \\
        \noalign{\smallskip}
        131007064 & 53.1107 & -27.7802 & 0.337 & 0.61$\pm$0.00 & 1.29$\pm$0.00 & 0.327$\pm$0.001 & 10.09$_{-0.11}^{+0.23}\hspace{1em}$ \\
        \noalign{\smallskip}
        133001029 & 53.1440 & -27.8137 & 0.215 & 3.37$\pm$0.06 & 2.93$\pm$0.03 & 0.762$\pm$0.003 & 10.58$_{-0.24}^{+0.00}\hspace{1em}$ \\
        \noalign{\smallskip}
        136002114 & 53.1166 & -27.7775 & 0.247 & 0.93$\pm$0.00 & 1.60$\pm$0.01 & 0.274$\pm$0.001 & 10.24$_{-0.17}^{+0.06}\hspace{1em}$ \\
        \noalign{\smallskip}
        139009211 & 53.1424 & -27.7651 & 0.366 & 0.66$\pm$0.01 & 1.53$\pm$0.01 & 0.544$\pm$0.004 & 9.01$_{-0.00}^{+0.40}\hspace{1em}$ \\
        \noalign{\smallskip}
        142002070 & 53.0847 & -27.7655 & 0.232 & 1.11$\pm$0.01 & 1.20$\pm$0.01 & 0.713$\pm$0.003 & 9.55$_{-0.19}^{+0.10}\hspace{1em}$ \\
        \noalign{\smallskip}
        \hline
        \end{tabular}
        \tablefoot{From left to right, the columns correspond to (1), (2), (3), and (4) the unique id, R.A., Decl. and redshift of the galaxy, taken from the emission line catalog of MUSE-Wide; (5), (6), and (7) the morphological parameters remeasured by GALFIT; (8) the stellar mass remeasured by matching the 3D-HST photometry catalog.}
\end{table*}

We remeasured the morphological parameters of these galaxies using HST F850LP images with the GALFIT software\footnote{\url{https://users.obs.carnegiescience.edu/peng/work/galfit/galfit.html}} \citep{Peng2010}. We used the remeasured $b/a$ and R$_{\rm e}$ derived by GALFIT as the final parameters. These morphological measurements are shown in Table \ref{table:morph_comparison}. The results measured by \cite{Wel2012} in F125W were also added in the table for reference. We note that the measured value of R$_{\rm e}$ of 133001029 is significantly higher than other measurements, but this is not important and has no effect on the following discussion and conclusions.

\begin{table}
        \caption{Comparison of morphological measurements\label{table:morph_comparison}}
        \centering
        \begin{tabular}{ccccccc}
        \hline\hline
                Unique ID & R$_{\rm e}$\tablefootmark{a} & R$_{\rm e}$\tablefootmark{b} & R$_{\rm e}$\tablefootmark{c} & b/a\tablefootmark{a} & b/a\tablefootmark{b} & b/a\tablefootmark{c} \\
        \hline
        102021103 & 0.75 & 0.76 & 0.78 & 0.68 & 0.66 & 0.68 \\
        106002013 & 0.64 & 0.68 & 0.62 & 1.00 & 0.87 & 0.82 \\
        113001007 & 1.52 & 1.79 & 1.38 & 0.45 & 0.32 & 0.31 \\
        122003050 & 1.84 & 1.73 & 1.44 & 0.57 & 0.58 & 0.57 \\
        124002008 & 1.19 & 1.11 & 1.07 & 0.42 & 0.32 & 0.34 \\
        131007064 & 0.68 & 0.61 & 0.52 & 0.33 & 0.33 & 0.32 \\
        133001029 & 1.64 & 3.37 & 1.73 & 0.75 & 0.76 & 0.74 \\
        136002114 & 0.99 & 0.93 & 0.77 & 0.41 & 0.27 & 0.27 \\
        139009211 & 0.88 & 0.66 & 0.47 & 0.45 & 0.54 & 0.46 \\
        142002070 & 1.35 & 1.11 & 1.19 & 0.63 & 0.71 & 0.63 \\
        \hline
        \end{tabular}
        \tablefoottext{a}{measured in the DESI Sweep catalog.}
        \tablefoottext{b}{measured in this work in F850LP, }
        \tablefoottext{c}{measured by \cite{Wel2012} in F125W.}
\end{table}

\subsection{Measurements from MUSE-Wide} \label{subsec:muse_measurement}

\subsubsection{Spectral fitting} \label{subsubsec:specfit}

We adopted a process similar  to that described by \cite{Yao_2022} in MaNGA to process these spectra. For each galaxy, we cut out a 40$\times$40 datacube from the field it belongs to and performed spectral fitting for each spaxel. Before the spectral fitting, we first corrected all spectra for Galactic extinction. We used the color excess \ebv map of the Milky Way \citep{Schlegel1998} and the extinction law presented by \cite{Cardelli1989}. We first moved each spectrum to the rest-frame, masked its main emission line, and then performed a low-order polynomial fitting to obtain the continua. After subtracting the continua from the origin spectra, we used MPFIT \citep{Markwardt2009} to perform Gaussian fitting on their emission lines (\ha, \hb, \oiii$\lambda$5007, and \nii$\lambda$6583) and then obtained their fluxes. The S/Ns of these emission lines wre estimated following \cite{Ly2014}.

\subsubsection{Spaxel binning} \label{subsubsec:binning}
We used the S/N of the emission line of the integrated spectrum of the entire galaxy to select the target galaxy. However, this cannot guarantee that the S/N of the emission line of each spaxel can meet the requirements for the accurate calculation of metallicity. Therefore, we stacked spaxels into bins to improve the S/N. We used the package VorBin\footnote{\url{https://pypi.org/project/vorbin/}}, which is a Python implementation of the 2D adaptive spatial binning method of \cite{Cappellari2003}. It uses Voronoi tessellations to bin data to a given minimum S/N.

VorBin requires the user to provide a value of the S/N of each spaxel. This average is greatly affected by low values and less affected by high values, which ensures that every emission line has a high enough S/N to calculate the metallicity. If we only input the S/N of one emission line (e.g., \oiii$\lambda$5007), the S/N of the other emission lines may not meet our requirements. Here, we defined a custom S/N, the harmonic mean of the S/N of the four emission lines (\ha, \hb, \oiii$\lambda$5007, and \nii$\lambda$6583), and used it as the input of VorBin.

\begin{table}
        \caption{Voronoi bins of the ten galaxies in our sample\label{table:bins_list}}
        \centering
        \begin{tabular}{ccccc}
        \hline\hline
        Unique ID &  & Bins & & Final bins\\
        \hline
        102021103 & $\hspace{2.5em}$ & 42 & $\hspace{2em}$ & 37 \\
        106002013 &  & 40 &  & 38 \\
        113001007 &  & 105 &  & 92 \\
        122003050 &  & 332 &  & 279 \\
        124002008 &  & 69 &  & 58 \\
        131007064 &  & 35 &  & 23 \\
        133001029 &  & 29 &  & 21 \\
        136002114 &  & 2 &  & 1 \\
        139009211 &  & 63 &  & 60 \\
        142002070 &  & 107 &  & 90 \\
        \hline
        \end{tabular}
        \tablefoot{The ``bins'' column represents the number of bins that is directly output by VorBin, and the ``final bins'' column represents the number of bins that we finally selected after the S/N cut.}
\end{table}

\begin{figure*}[ht!]
        \centering
        \includegraphics[width=1\textwidth]{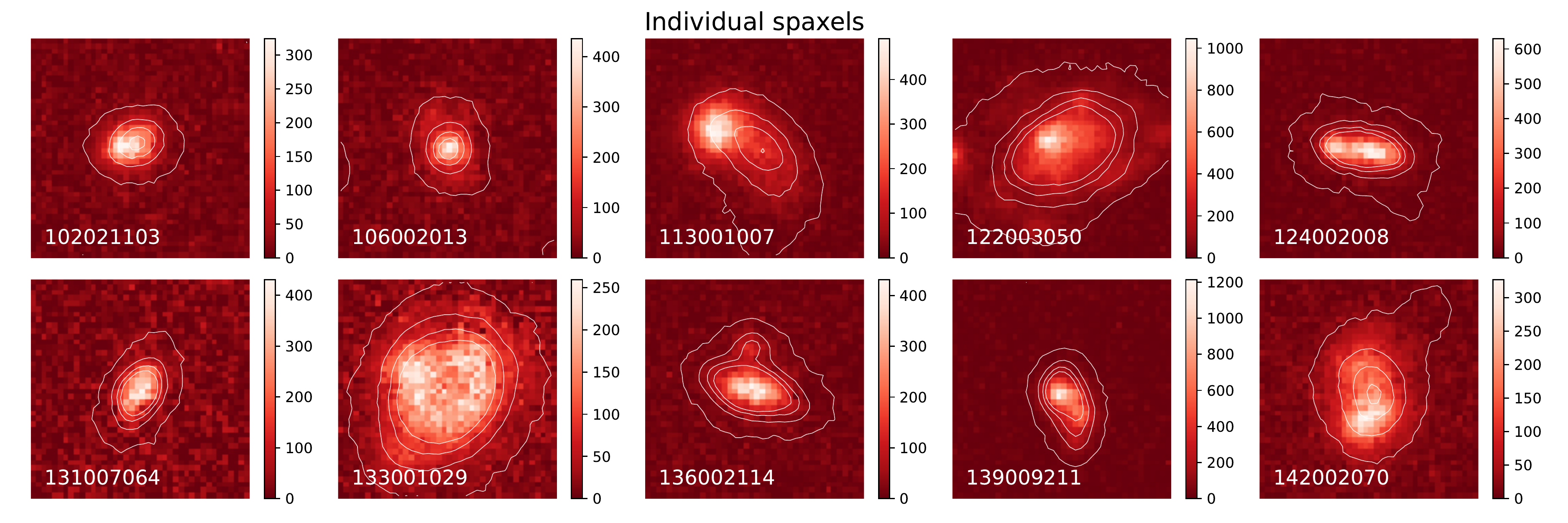}
        \includegraphics[width=1\textwidth]{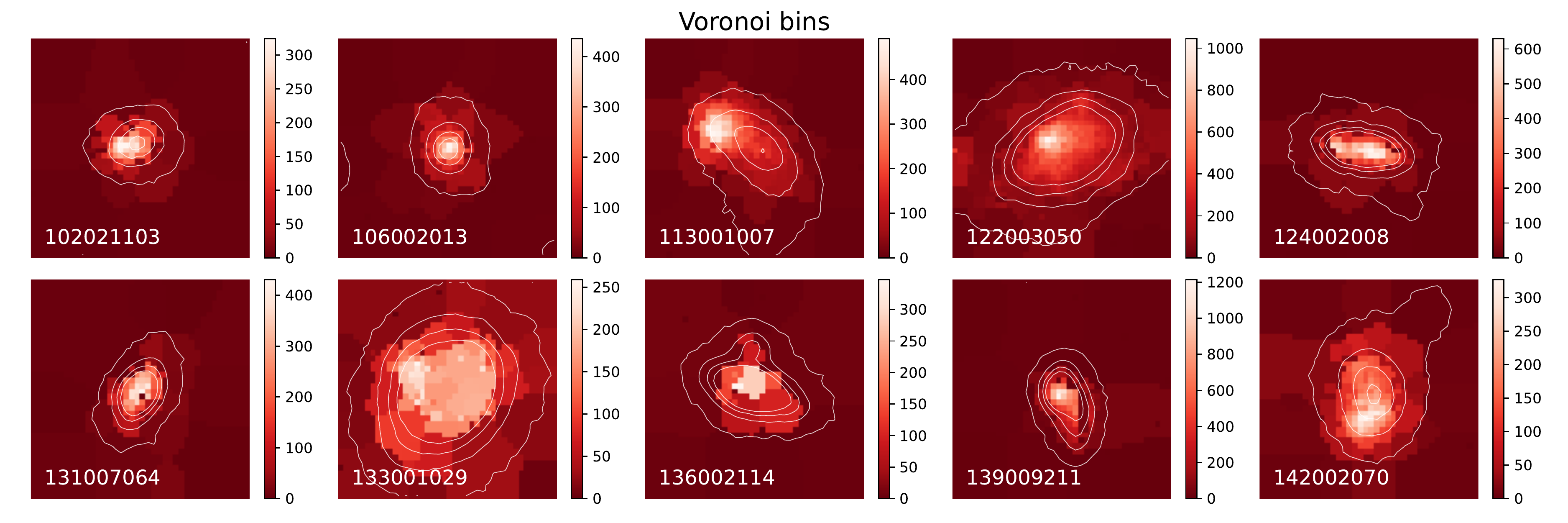}
        \caption{\ha{} emission maps without dust extinction correction of our ten galaxies. The top two rows are before VorBin, and the bottom two rows are after VorBin. The color bars indicate the flux of \ha{} in units of 10$^{-20}$ erg s$^{-1}$ cm$^{-2}$. White contours represent the isophotes of their MUSE white-light images.}
        \label{fig:ha_map}
\end{figure*}

We stacked the spectra of all spaxels belonging to each bin and then averaged the stacked spectrum. We used the same steps as in Sect. \ref{subsubsec:specfit} to refit the averaged spectrum in each bin. We discarded the bins whose custom S/N was still lower than 3 after stacking. Finally, we obtained 699 Voronoi bins in our ten sample galaxies. Table \ref{table:bins_list} lists the final number of Voronoi bins for each galaxy, and Fig. \ref{fig:ha_map} shows the \ha{} emission map before and after VorBin. We corrected the fluxes of emission lines for dust extinction by comparing the Balmer decrement (e.g., H$\alpha$/H$\beta$ ratio) to the intrinsic values under the case B assumption. The Balmer decrement is insensitive to the temperature and electron density, and here we adopted an intrinsic flux ratio of (H$\alpha$/H$\beta$)$_0$=2.86 \citep{Hummer1987}. We used the \cite{Calzetti2000} dust-attenuation law to derive the color excesses \ebv and correct for the dust extinction for the emission line fluxes. When a spectrum has an observed value of F(H$\alpha$)/F(H$\beta$)$<$2.86, we regarded it as zero extinction. We used the BPT diagram \citep{Baldwin1981} and found that all bins are below the curve of \cite{Kewley2001}. We measured their equivalent width of \ha{} and found that all are higher than 6\AA, confirming that all bins are dominated by star formation.

\subsubsection{SFR surface density} \label{subsubsec:sfr}

We used the H$\alpha$ luminosity to determine the dust-corrected SFR for each spaxel. The SFR calibration from \cite{Kennicutt1998} is
\begin{equation}\label{eq:sfr_calc}
        {\rm SFR(M_{\odot}~yr^{-1})}=4.4 \times 10^{-42} L_{\mathrm{H}\alpha}\mathrm{(erg~s^{-1})}.
\end{equation}
Projection correction is required to calculate the surface density of each property. The inclination angle is derived using
\begin{equation}
        \cos^2{i}=\frac{(b/a)^2-q^2}{1-q^2},
\end{equation}
where $i$ is the inclination angle, $q$ is the intrinsic axis ratio, and $b/a$ is from GALFIT. We adopted $q$=0.13 \citep{Giovanelli1994} for our analysis.
The size of each spaxel of the MUSE wide-field mode is $0\farcs2 \times 0\farcs2$, so the projection corrected area is given by
\begin{equation}
        A=\left[D(z) \times 0.2 \times \frac{\pi}{3600 \times 180}\right]^2 \times \frac{1}{\cos{i}},
\end{equation}
where $D(z)$ is the angular diameter distance of the galaxy. As a result, the $\Sigma_{\rm SFR}$ of each spaxel is ${\rm \Sigma_{\rm SFR} = SFR/}A$.

\subsubsection{Oxygen abundance} \label{subsubsec:metal}

The gas-phase metallicity is usually described by the oxygen abundance (12+log(O/H)). The most reliable approach to derive the oxygen abundance is the electron temperature (T$_{\rm e}$) method, which is based on the ratios of the faint auroral-to-nebular emission lines \citep{Lin2017}. However, the \oiii$\lambda$4363 auroral line is too weak to be observed. Only a few studies \citep{Yao_2022} have used the low-redshift IFS survey, let alone the high redshift. Therefore, we used the strong-line calibration relation to calculate the abundance. Commonly used calibration relations include R23 \citep{Kobulnicky2004}, N2O2 \citep{Kewley2002}, N2, and O3N2 \citep{Pettini2004}. Here, according to our sample selection conditions, the O3N2 method was adopted. The diagnostic O3N2 index is defined as
\begin{equation}
        {\rm O3N2}\equiv\log{\left(\frac{\oiii\lambda5007}{\rm H\beta}\times\frac{\rm H\alpha}{\nii\lambda6583}\right)},
\end{equation}
and the calibration of O3N2 diagnostic we used, improved by \cite{Marino2013} using CALIFA data, is
\begin{equation}
        12+\log{\rm (O/H)}=8.505-0.221 \times {\rm O3N2},
\end{equation}
with O3N2 ranging from -1.1 to 1.7. The typical error for the metallicity calibration with the O3N2 diagnostic is 0.08 dex.

\subsection{Measurements from 3D-HST} \label{subsec:3dhst_measurement}
Although the IFS of MUSE-Wide can provide spatially resolved continua, their S/Ns are too low to determine their stellar mass well, and the wavelength range is limited. Therefore, we used multiband photometry data from HST.

\subsubsection{Image processing of HST}

The PSF and the sampling rate of the image provided by the 3D-HST are not the same as those from MUSE. In order to match each pixel of the HST image with MUSE, we used the photutils\footnote{\url{https://github.com/astropy/photutils}} \citep{Bradley2021} package to create a convolved kernel from HST to MUSE. The 3D-HST data release provides the PSFs of HST. The PSFs in different fields of MUSE were taken from Table 2 of \cite{Urrutia2019} and were assumed to be Gaussian. Then we used the reproject\footnote{\url{https://github.com/astropy/reproject}} package to project the HST image to the exact coordinates and sampling rate as MUSE.

For the matched HST images, we stacked the electron flux and error (in units of electrons s$^{-1}$) in each bin and used the inverse sensitivity (in units of erg cm$^{-2}$ \AA$^{-1}$ electron$^{-1}$) parameter given by the 3D-HST to convert the electrons s$^{-1}$ into AB magnitudes to facilitate subsequent processing.

\subsubsection{Mass surface density}

We derived stellar mass surface densities ($\Sigma_*$) by measuring the photometry in multiple HST bands (F435W, F606W, F775W, F850LP, F125W, F140W, and F160W) for each bin corresponding to MUSE. We used the FAST \citep{Kriek2009} SED-fitting code, with the \cite{Bruzual2003} stellar synthesis models, an exponentially decaying star-forming history, and the \cite{Calzetti2000} dust attenuation law. We converted the output masses into mass surface densities by $\Sigma_*=M_*/A$.

\section{Result and discussion} \label{sec:result}

This section mainly lists the results of the rMZR of our sample, and we also give the results of the rSFMS and metallicity gradients and compare them with other works. The IMF and metallicity calibrator is consistent. 

\subsection{ $\Sigma_*$-$\Sigma_{\rm SFR}$ relation (rSFMS)} \label{subsec:rsfms}

\begin{figure}[ht!]
        \centering
        \includegraphics[width=0.9\columnwidth]{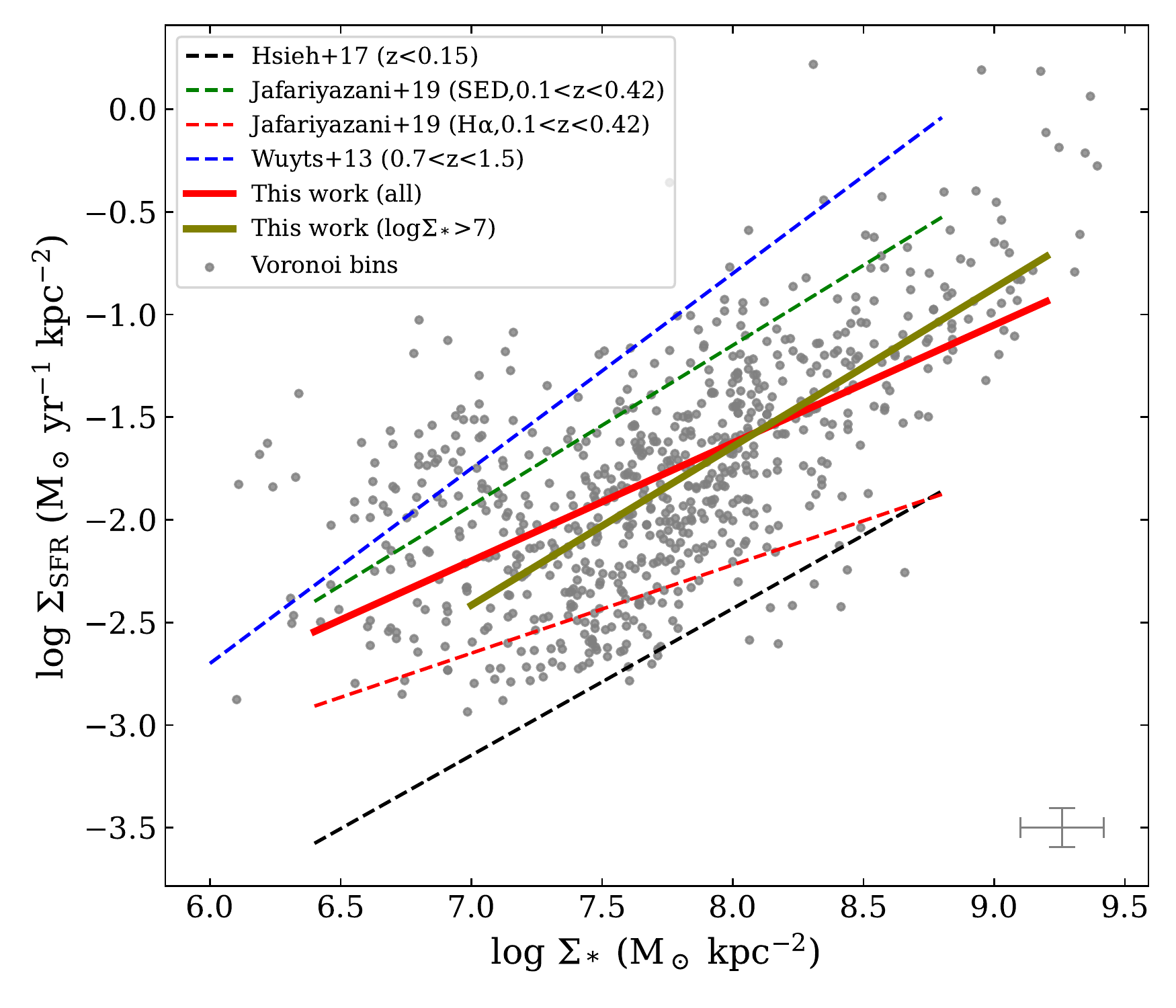}
        \caption{rSFMS of our sample compared with other works. The gray dots represent our Voronoi bins, and the error bar in the lower right corner represents the average value of the uncertainty on each axis. The thick red line represents the linear fit of all Voronoi bins, and the thick green line only fits bins whose $\log{\Sigma_*}\geqslant$7. The dashed black line comes from \cite{Hsieh2017}, the dashed blue line comes from \cite{Wuyts2013}, and the dashed red and green lines come from the H$\alpha$ SFR and from the SED-fitting SFR of \cite{Jafariyazani2019}, respectively. }
        \label{fig:sigma_sfr}
\end{figure}

The results of the rSFMS are shown in Fig. \ref{fig:sigma_sfr}. The average uncertainties of $\log{\Sigma_*}$ and $\log{\Sigma_{\rm SFR}}$ are 0.32 and 0.19 dex, respectively. They are displayed as error bars in the lower right corner of Fig. 4. Our SFR is located between that of the local universe \citep[$z<0.15$;][]{Hsieh2017} and that of the universe with higher redshift \citep[$0.7<z<1.5$;][]{Wuyts2013}, which is in line with the expectation that the SFR will increase with the increase in redshift. We performed an ordinary least-squares linear fitting on our sample, and the result is
\begin{equation}
        \log{\Sigma_{\rm SFR}}=(0.571\pm0.026)\log{\Sigma_*}-(6.190\pm0.203),
\end{equation}
whose slope is much shallower than those of \cite{Wuyts2013} (0.95), \cite{CanoDiaz2016}, (0.72) and \cite{Hsieh2017} (0.715), whether it is compared with that at high or at low redshift.

We also compared the results with \cite{Jafariyazani2019}, who used H$\alpha$ and SED-fitting to derive the SFR based on the same MUSE-Wide data. The fitted slopes are 0.43 and 0.78, respectively. We note that the slope they obtained with H$\alpha$ is shallower than ours and that of their SED-fitting. They suggested that the slope of H$\alpha$ is shallower than that of the SED-fitting because of the inside-out quenching of star formation. However, because in our sample the rSFMS shape flattens at the low-mass end, the S/N selection might lead to a shallow slope of our H$\alpha$-based rSFMS. We estimated the completeness of the low-mass end ($\log{\Sigma_*}<$7.0) of our sample and find that it is lower than 50\%. Therefore, we refit the bins with $\log{\Sigma_*}\geqslant$7.0, and the result is
\begin{equation}
        \log{\Sigma_{\rm SFR}}=(0.771\pm0.032)\log{\Sigma_*}-(7.812\pm0.249),
\end{equation} 
whose slope is between that of the local universe and that of the higher-redshift universe. The slope is consistent with the simulation of \cite{Trayford2019}, who reported that the slope of the rSFMS grows with increasing redshift.

\subsection{ $\Sigma_*$-$Z$ relation (rMZR)} \label{subsec:rmzr}

\begin{figure}[ht!]
        \centering
        \includegraphics[width=0.9\columnwidth]{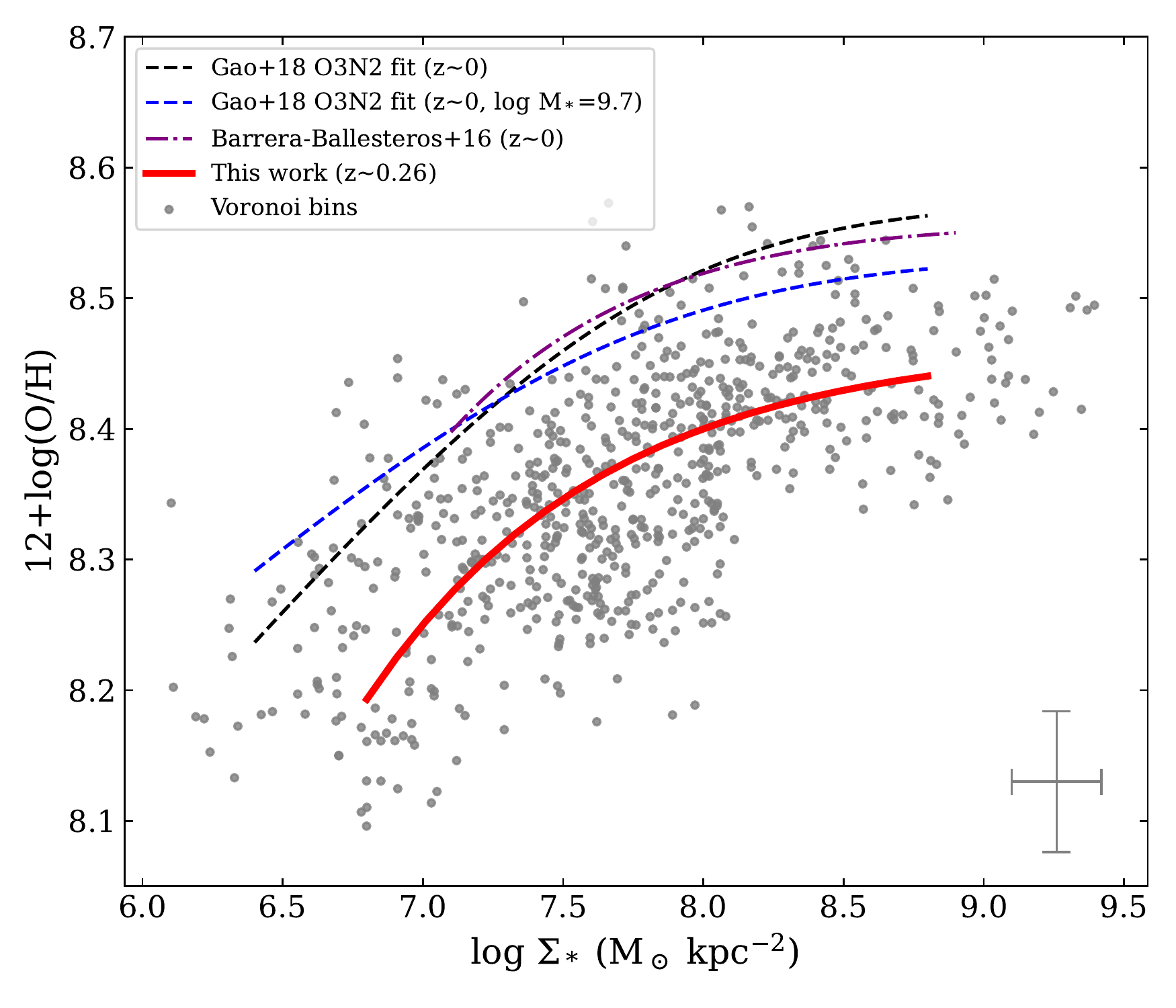}
        \caption{rMZR of our sample compared with that of other works. The gray dots represent our bins, and the error bar in the lower right corner represents the average value of the uncertainty on each axis. The thick red curve represents the fit of all gray crosses. The dashed black curve comes from \cite{Gao2018b}, the dashed blue curve is the fit from \cite{Gao2018b} at a fixed $\log{M_*}$=9.7, and the dash-dotted purple curve comes from \cite{BarreraBallesteros2016}. The dash-dotted purple curve is shifted by 0.24 dex to the left to correct for the systematic error caused by the IMF \citep{Sanchez2016}.}
        \label{fig:sigma_z}
\end{figure}

Fig. \ref{fig:sigma_z} shows the results of rMZR. The average uncertainty of metallicity derived from the noise is $\sim$0.11 dex, displayed as the vertical error bar in the lower right corner. We performed an orthogonal distance regression (ODR) fitting to minimize the sum of the squares of both the $x$ residual and the $y$ residual to our sample, and the fitting function is from \cite{Moustakas2011},\begin{equation}
        12+\log{({\rm O/H})}=a+b(\log{\Sigma_*}-c)e^{-(x-c)}.
\end{equation} 
The coefficients corresponding to the ODR fit of the above function to our Voronoi bins are $a$=8.47$\pm$0.02, $b$=0.0002$\pm$0.0011, and $c$=12.36$\pm$5.08. The residual variance is 0.006.

Here we mainly compare our result with the local star-forming galaxies selected from MaNGA DR14 \citep{Gao2018b}. Taking into account the systematic errors between different strong-line metallicity calibrators, we used the O3N2 calibrator \citep{Marino2013}, which is also adopted in \cite{Gao2018b}. Fig. \ref{fig:sigma_z} shows that our high-redshift sample shows an rMZR similar to that of the local galaxies, but our metallicities are below those of the local galaxies overall. The average downward shift is $\sim$0.11 dex. In addition, considering that $M_*$ also has a certain influence on the rMZR \citep{BarreraBallesteros2016,Gao2018b}, in order to rule out whether our low metallicity is caused by low $M_*$ (our average log $M_*$ is $\simeq$9.7), we used the dashed blue curve in Fig. \ref{fig:sigma_z} to indicate the $M_*$-$\Sigma_*$-Z relation fitted by \cite{Gao2018b} at the same average $M_*$. When we consider their $ M_*$, the average downward shift is $\sim$0.09 dex. Therefore, we can conclude that the low local metallicity of our sample is not caused by their low $M_*$, but that the local metallicity evolves with redshift. This result is consistent with the scenario that the length of the history of star formation determines the local metallicity. When they reach the same $\Sigma_*$, galaxies with higher redshifts have a shorter history of star formation, a lower metal production, and a higher SFR and more intense gas outflow than lower redshifts galaxies.

\begin{figure}[ht!]
        \centering
        \includegraphics[width=0.9\columnwidth]{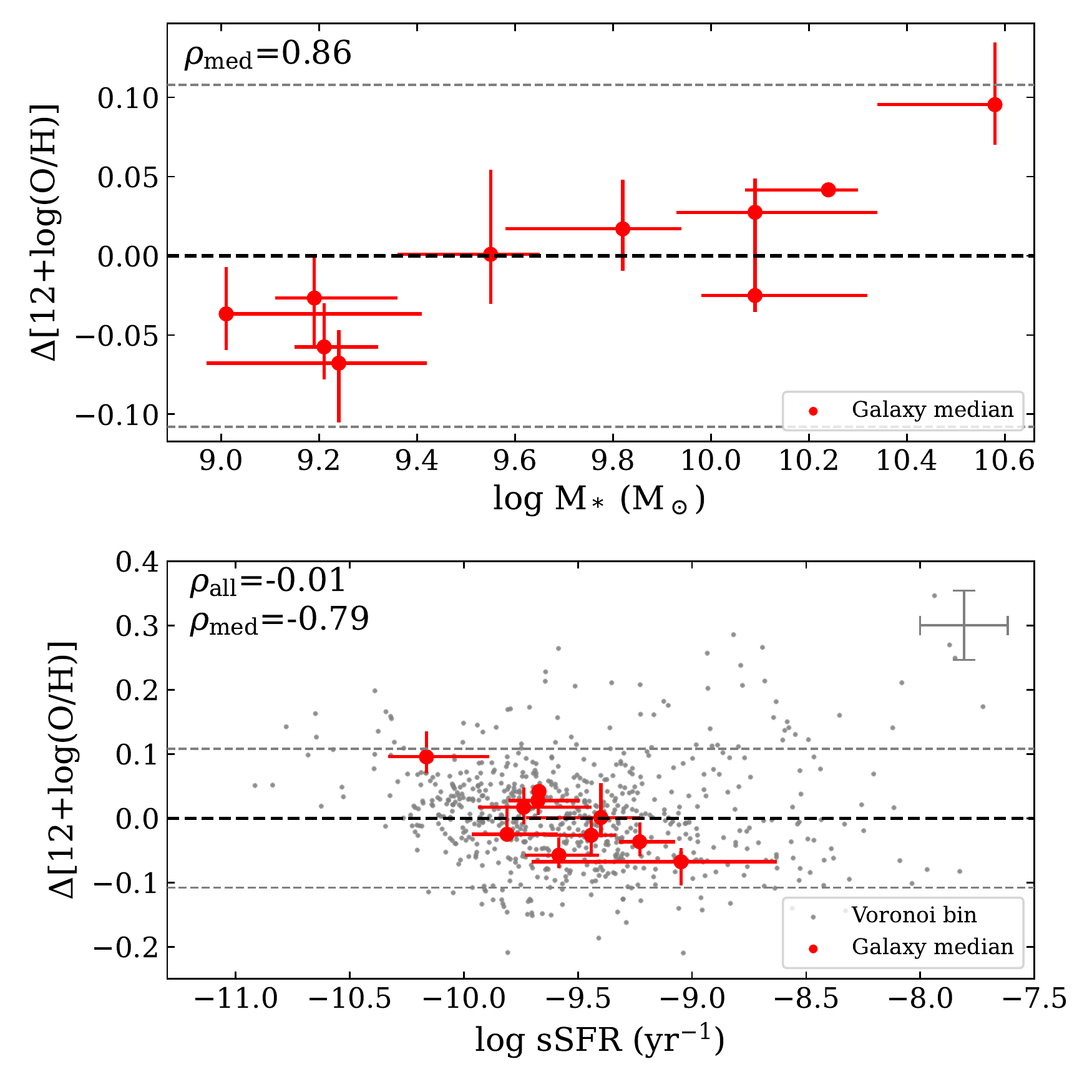}
        \caption{Relation between $\Delta$[12+log(O/H)] and sSFR and $M_*$. Each red dot represents the median value of a galaxy, whose error bar indicates the 32\% and 68\% quantiles of the distribution of all Voronoi bins of the galaxy. The gray dots in the lower panel represent Voronoi bins, and the error bar in the upper right corner indicates the uncertainty on each axis. The horizontal dashed gray line represents the average value of the uncertainty of our measured metallicity. The upper left corner is the Pearson correlation coefficient.}
        \label{fig:res_ssfr}
\end{figure}

In addition, we also compared the relation between the residuals of our fitting and other properties. \cite{BarreraBallesteros2016} pointed out that the residuals in the best-fitting rMZR correlated with specific star formation rate (sSFR=$\Sigma_{\rm SFR}$/$\Sigma_*$) and $M_*$. However, \cite{Gao2018b} directly reported the relation between $M_*$, $\Sigma_*$, and Z. In order to verify whether these relations still exist at higher redshift, we show the relation between the residual of metallicity ($\Delta$[12+log(O/H)]=Z$_{\rm obs}$-Z$_{\rm fit}$) and sSFR and $M_*$ in Fig. \ref{fig:res_ssfr}.

Similar to \cite{BarreraBallesteros2016}, we find that these two properties have very little influence on the residuals of the fitting of rMZR. The residual and $M_*$ have an apparent positive correlation with a Pearson correlation ($\rho$) coefficient of $\sim$0.86. However, the correlation between the residual and sSFR is very weak ($\rho \sim$-0.01), and the scatter is much more significant when the Voronoi bin is considered separately. A large uncertainty in the measurement probably causes this. When we consider the median value of each galaxy, the correlation is negative, with $\rho$  $\sim$-0.79, but it is still weaker than $M_*$. This result means that $M_*$ is still an excellent third parameter in the rMZR at $z\sim$0.26, rather than sSFR. However, it is also necessary to indicate that although the correlation is strong after taking the median for each galaxy, the influence of these two parameters is within the uncertainty derived from the noise ($\sim$0.11 dex) and the calibrator ($\sim$0.08 dex) of the metallicity. That is, the uncertainty of the metallicity measurement is the primary source of the dispersion in the rMZR at $z\sim$0.26 derived by us.

\subsection{Metallicity gradient} \label{subsec:gradient}

\begin{figure}[ht!]
        \centering
        \includegraphics[width=0.9\columnwidth]{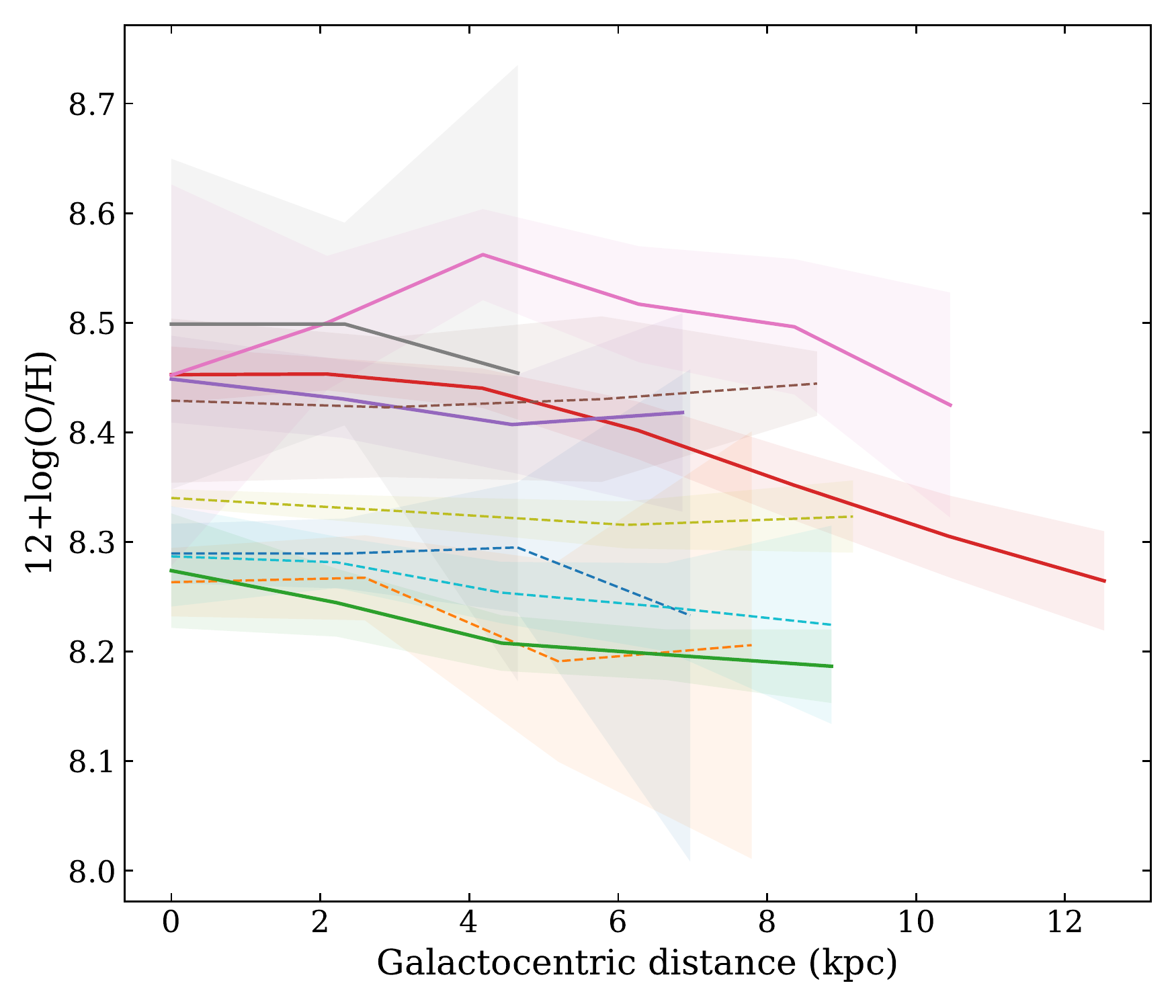}
        \caption{Metallicity profiles along the galactocentric radius of our selected galaxies. The semitransparent filled area indicates the metallicity error. Solid lines are galaxies whose R$_{\rm e}>$ FWHM, and dashed lines are galaxies whose R$_{\rm e}\leqslant$ FWHM.}
        \label{fig:metal_profile}
\end{figure}

We also measured the profiles in the metallicity of these galaxies along their radius. We divided each of our galaxies into annuli according to its ellipticity and position angle. All spectra in the annulus were stacked and averaged, and then the same fitting process as Sect. \ref{subsubsec:specfit} was performed again. The width of the rings is $\sim0\farcs6$. The PSF also has an effect on the measured value of metallicity gradient, which appears to flatten the gradient, especially in galaxies with a low R$_{\rm e}$/FWHM \citep{Belfiore2017,Acharyya2020}. In our sample, only five galaxies (113001007, 122003050, 124002008, 133001029, and 136002114) have R$_{\rm e}$ greater than the FWHM of their PSFs. Fig. \ref{fig:metal_profile} shows the metallicity profiles along the radius. The R$_{\rm e}\leqslant$FWHM of the other five galaxies is also plotted in this figure. 

We performed a linear fit to the metallicity profiles of our galaxies to obtain their gradients. The results of our five galaxies whose R$_{\rm e}$ is larger than FWHM are -0.009$\pm$0.002, -0.015$\pm$0.002, -0.007$\pm$0.002, -0.007$\pm$0.006, and -0.006$\pm$0.007, respectively, and the units are all dex/kpc. In the figure and the fitting results, the metallicity gradients of our galaxies all have a weak negative trend. Nevertheless, these slopes are shallower than those of local galaxies, consistent with other studies of metallicity gradients at similar redshifts \citep[e.g.,][]{Carton2018}.

\subsection{Is the rMZR at $z\sim$0.26 an extension of rFMR under high SFR?} \label{subsec:rFMR}

\begin{figure}[ht!]
        \centering
        \includegraphics[width=0.9\columnwidth]{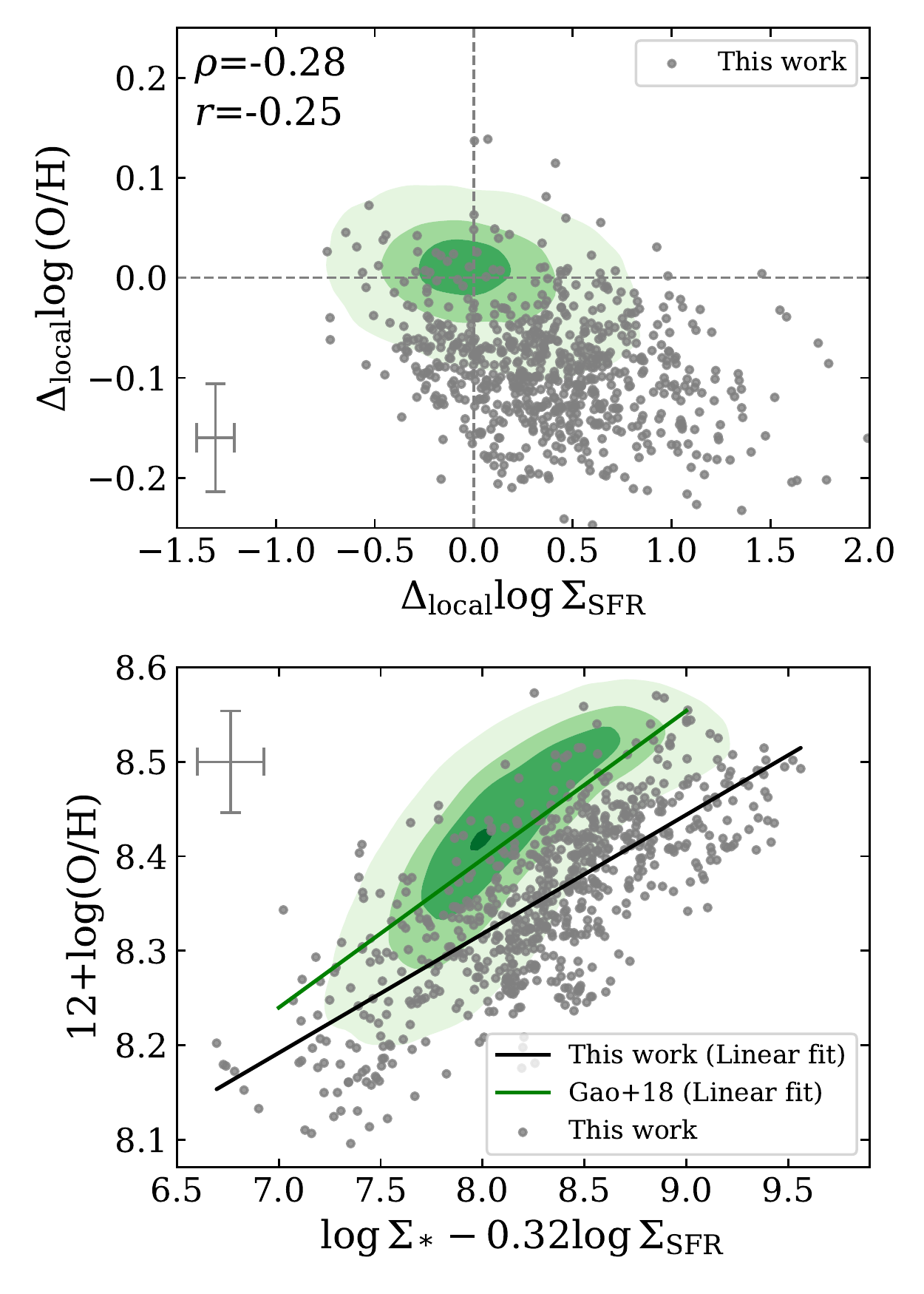}
        \caption{Residuals of the rMZR and residuals of rSFMS fitted by \cite{Gao2018b} (top panel).  rFMR diagram assuming $\alpha$=0.32 (bottom panel). The two lines with different colors represent the results of linear fitting of the two rFMRs. The gray crosses represent our Voronoi bins and their uncertainties, and the contour filled with green represents the sample of \cite{Gao2018b} in each panel.}
        \label{fig:fmr}
\end{figure}

Some studies have postulated that the FMR will not evolve with redshift during the last half of cosmic history \citep[e.g.,][]{Mannucci2010,Cresci2012,Huang2019}. The SFR of a galaxy will increase as the redshift increases. The decrease in metallicity at higher redshift extends the FMR under high SFR. At the same time, some other studies questioned the existence of the (r)FMR or suggested that the SFR should be substituted by another parameter \citep[e.g.,][]{Sanchez2013,Bothwell2016,BarreraBallesteros2018}. Compared with the local ones, our results have higher $\Sigma_{\rm SFR}$ and lower metallicity. In order to verify whether the rFMR exists in our sample and whether our result is an extension of the rFMR under high $\Sigma_{\rm SFR}$, we compared our sample with that \cite{Gao2018b}, shown in Fig. \ref{fig:fmr}. The SFR coefficient ($\alpha$) of FMR we adopted is 0.32 \citep{Mannucci2010}. The $\Delta_{\rm local}$ means the difference between the observed value and the value predicted by the fitted local scale relation in \cite{Gao2018b} (observed value minus predicted value).

The top panel of Fig. \ref{fig:fmr} shows that our Voronoi bins at $z\sim$0.26 are all located in the bottom right corner of the local sample, and there is no apparent correlation between the residuals of $\Sigma_{\rm SFR}$ and the metallicity residuals. The Pearson correlation coefficient is $\rho$=-0.28 with $p\simeq6\times10^{-14}$ , and the Spearman correlation coefficient is $r$=-0.25 with $p\simeq3\times10^{-11}$. These coefficients indicate a weak correlation. As a result, we cannot conclude that our sample has a significant rFMR, which is consistent with the research results for the local universe. The deficiency of metal and the intensity of the star formation in our galaxies could be a measurable effect of galaxy evolution.

The bottom panel of Fig. \ref{fig:fmr} directly shows the rFMR of our Voronoi bins at $z\sim$0.26 and the local sample. We find that when $\alpha$=0.32, the rFMR of our sample does not entirely coincide with the local rFMR. The $\alpha$ derived from different metallicity calculation methods may be different \citep[e.g.,][$\alpha$=0.66]{Andrews2013}. We tried to determine $\alpha$ by minimizing the dispersion of the fit of rFMR, and we found that when $\alpha\simeq$0.51, the dispersion of the rFMR reaches a minimum value of $\sim$0.061 dex, but it is only $\sim$0.007 dex lower than when $\alpha$=0 (i.e., there is no rFMR at all). This improvement is much smaller than that of \cite{Mannucci2010}. However, even if $\alpha$ increases to 0.51, our sample still does not coincide with the local sample of \cite{Gao2018b} in the rFMR diagram. This result means that our sample is not an extension of the rMZR under high $\Sigma_{\rm SFR}$ if rFMR exists and $\alpha$ is not much different from the global ones. Therefore, the existence of rFMR and its evolution with redshift need to be studied further.

\section{Summary} \label{sec:conclusion}

We used the Legacy Surveys to select sources in the fields of MUSE-Wide DR1 and selected ten emission line galaxies dominated by star formation with an average redshift of $z\sim$0.26. We measured their $\Sigma_*$ with the 3D-HST images and derived their spatially resolved metallicity and $\Sigma_{\rm SFR}$ with the MUSE-Wide IFS data. We summarize the main conclusions below.

\begin{enumerate}
        \item The rMZR exists at $z\sim$0.26 and the shape is similar to that of the local universe, but the average metallicity is 0.11 dex lower than that of the local universe. Both $M_*$ and sSFR have weak effects on the residuals of rMZR, and the effects are weaker than the uncertainties in metallicity measurements. After considering the influence of $M_*$, the decrease in metallicity is still 0.09 dex. This result indicates that the rMZR evolves with redshift.
        
        \item After we removed the low-$\Sigma_*$ parts that may cause serious selection effects, our slope of the rSFMS at $z\sim$0.26 is $\sim$0.771, which is between the slopes of the local universe and the higher-redshift universe, which is consistent with the trend predicted by simulations.
        
        \item The metallicity gradient is negative but low, within the range of other the gradients in previous work. However, due to the error and PSF, the measurement result of the gradient is not accurate. 
        
        \item We have no clear evidence that the rFMR exists at $z\sim$0.26. Moreover, the high $\Sigma_{\rm SFR}$ and low metallicity we observed are not the extension of the local rFMR under high $\Sigma_{\rm SFR}$. The evolution of the global properties of galaxies with redshift may be an integration effect of the evolution of the spatially resolved properties of galaxies.
\end{enumerate}

However, the sample we used is small, the redshift range is narrow, and there is a significant measurement error. The questions remain how the rMZR evolves with the redshift and how it reproduces the evolution of the MZR with redshift. We look forward to a broader and deeper IFS survey in multiple bands to answer these questions.

\begin{acknowledgements}
This work is supported by the Strategic Priority Research Program of Chinese Academy of Sciences (No. XDB 41000000), the National Key R\&D Program of China (2017YFA0402600, 2017YFA0402702), the NSFC grant (Nos. 11973038 and 11973039), and the Chinese Space Station Telescope (CSST) Project. HXZ also thanks a support from the CAS Pioneer Hundred Talents Program.

This work is based on observations taken by the MUSE-Wide Survey as part of the MUSE Consortium.

This work is based on observations taken by the 3D-HST Treasury Program (HST-GO-12177 and HST-GO-12328) with the NASA/ESA Hubble Space Telescope, which is operated by the Association of Universities for Research in Astronomy, Inc., under NASA contract NAS5-26555.

The Legacy Surveys consist of three individual and complementary projects: the Dark Energy Camera Legacy Survey (DECaLS; Proposal ID \#2014B-0404; PIs: David Schlegel and Arjun Dey), the Beijing-Arizona Sky Survey (BASS; NOAO Prop. ID \#2015A-0801; PIs: Zhou Xu and Xiaohui Fan), and the Mayall z-band Legacy Survey (MzLS; Prop. ID \#2016A-0453; PI: Arjun Dey). DECaLS, BASS and MzLS together include data obtained, respectively, at the Blanco telescope, Cerro Tololo Inter-American Observatory, NSF’s NOIRLab; the Bok telescope, Steward Observatory, University of Arizona; and the Mayall telescope, Kitt Peak National Observatory, NOIRLab. The Legacy Surveys project is honored to be permitted to conduct astronomical research on Iolkam Du’ag (Kitt Peak), a mountain with particular significance to the Tohono O’odham Nation.

NOIRLab is operated by the Association of Universities for Research in Astronomy (AURA) under a cooperative agreement with the National Science Foundation.

This project used data obtained with the Dark Energy Camera (DECam), which was constructed by the Dark Energy Survey (DES) collaboration. Funding for the DES Projects has been provided by the U.S. Department of Energy, the U.S. National Science Foundation, the Ministry of Science and Education of Spain, the Science and Technology Facilities Council of the United Kingdom, the Higher Education Funding Council for England, the National Center for Supercomputing Applications at the University of Illinois at Urbana-Champaign, the Kavli Institute of Cosmological Physics at the University of Chicago, Center for Cosmology and Astro-Particle Physics at the Ohio State University, the Mitchell Institute for Fundamental Physics and Astronomy at Texas A\&M University, Financiadora de Estudos e Projetos, Fundacao Carlos Chagas Filho de Amparo, Financiadora de Estudos e Projetos, Fundacao Carlos Chagas Filho de Amparo a Pesquisa do Estado do Rio de Janeiro, Conselho Nacional de Desenvolvimento Cientifico e Tecnologico and the Ministerio da Ciencia, Tecnologia e Inovacao, the Deutsche Forschungsgemeinschaft and the Collaborating Institutions in the Dark Energy Survey. The Collaborating Institutions are Argonne National Laboratory, the University of California at Santa Cruz, the University of Cambridge, Centro de Investigaciones Energeticas, Medioambientales y Tecnologicas-Madrid, the University of Chicago, University College London, the DES-Brazil Consortium, the University of Edinburgh, the Eidgenossische Technische Hochschule (ETH) Zurich, Fermi National Accelerator Laboratory, the University of Illinois at Urbana-Champaign, the Institut de Ciencies de l’Espai (IEEC/CSIC), the Institut de Fisica d’Altes Energies, Lawrence Berkeley National Laboratory, the Ludwig Maximilians Universitat Munchen and the associated Excellence Cluster Universe, the University of Michigan, NSF’s NOIRLab, the University of Nottingham, the Ohio State University, the University of Pennsylvania, the University of Portsmouth, SLAC National Accelerator Laboratory, Stanford University, the University of Sussex, and Texas A\&M University.

BASS is a key project of the Telescope Access Program (TAP), which has been funded by the National Astronomical Observatories of China, the Chinese Academy of Sciences (the Strategic Priority Research Program “The Emergence of Cosmological Structures” Grant \# XDB09000000), and the Special Fund for Astronomy from the Ministry of Finance. The BASS is also supported by the External Cooperation Program of Chinese Academy of Sciences (Grant \# 114A11KYSB20160057), and Chinese National Natural Science Foundation (Grant \# 11433005).

The Legacy Survey team makes use of data products from the Near-Earth Object Wide-field Infrared Survey Explorer (NEOWISE), which is a project of the Jet Propulsion Laboratory/California Institute of Technology. NEOWISE is funded by the National Aeronautics and Space Administration.

The Legacy Surveys imaging of the DESI footprint is supported by the Director, Office of Science, Office of High Energy Physics of the U.S. Department of Energy under Contract No. DE-AC02-05CH1123, by the National Energy Research Scientific Computing Center, a DOE Office of Science User Facility under the same contract; and by the U.S. National Science Foundation, Division of Astronomical Sciences under Contract No. AST-0950945 to NOAO.

This research made use of Photutils, an Astropy package for detection and photometry of astronomical sources \cite{Bradley2021}.
\end{acknowledgements}

%
%

\bibliographystyle{aa} 
\bibliography{aanda.bib} 

\end{document}